\documentclass[pra,twocolumn,preprintnumbers,superscriptaddress,amsmath,amssymb]{revtex4}

\usepackage{amssymb}
\usepackage{amsmath}
\usepackage{amsfonts}
\usepackage{wasysym}
\usepackage{graphicx}
\usepackage{dcolumn}
\usepackage{bm}
\usepackage{color}
\usepackage[colorlinks,linkcolor=blue, urlcolor=blue, anchorcolor=blue, citecolor=blue]{hyperref}

\begin{document}

\title{Low-energy scatterings and pseudopotential of polarized quadrupoles}

\author{Fulin Deng}
\affiliation{Key Laboratory of Artificial Micro- and Nano-structures of Ministry of Education, School of Physics and Technology, Wuhan University, Wuhan, Hubei 430072, China}
\affiliation{CAS Key Laboratory of Theoretical Physics, Institute of Theoretical Physics, Chinese Academy of Sciences, Beijing 100190, China}

\author{Wenxian Zhang}
\affiliation{Key Laboratory of Artificial Micro- and Nano-structures of Ministry of Education, School of Physics and Technology, Wuhan University, Wuhan, Hubei 430072, China}
\affiliation{Wuhan Institute of Quantum Technology, Wuhan, Hubei 430206, China}

\author{Su Yi}
\email{syi@itp.ac.cn}
\affiliation{CAS Key Laboratory of Theoretical Physics, Institute of Theoretical Physics, Chinese Academy of Sciences, Beijing 100190, China}
\affiliation{CAS Center for Excellence in Topological Quantum Computation \& School of Physical Sciences, University of Chinese Academy of Sciences, Beijing 100049, China}
\affiliation{Peng Huanwu Collaborative Center for Research and Education, Beihang University, Beijing 100191, China}

\begin{abstract}
We investigate the low-energy scattering properties of two identical particles interacting via the polarized quadrupolar interaction. It is shown that a series of $s$- and $p$-wave resonances appear for identical bosons and fermions, respectively, as the strength of the quadrupolar interaction increases. Interestingly, scattering resonances also appear on the generalized scattering length corresponding to the coupling between the $s$ and $d$ waves. This observation inspires us to propose a new pseudopotential for the quadupolar interaction. We also explore the bound-state properties of two particles in both free space and harmonic traps.
\end{abstract}

\maketitle

\section{Introduction}

Interatomic interactions in ultracold atomic gases are of fundamental importance in determining the properties of the systems. Under a ultralow temperature, the van der Waals force between two neutral atoms can be described by a contact potential characterized by a single $s$-wave scattering length. Such a simplification has gained great success in cold atomic physics (see, e.g., Refs.~\cite{Stringari1999,Bloch2008}). In addition to the isotropic contact interaction, the long-range and anisotropic dipole-dipole interaction may become dominant and lead to the dipolar quantum gases for particles with large magnetic moments~~\cite{Baranov2002,Baranov2008,Lahaye2009,Baranov2012,Bottcher2021,Guo2021}. For the mean-field treatment of the dipolar gases, an important step is to identify that the interaction between two identical bosonic dipoles can be modeled by a pseudopotential consisting of a $s$-wave contact potential and a bare dipole-dipole interaction potential whose scattering amplitude reproduces that of the real potential away from scattering resonance~\cite{You1998,DDI-pseu1,DDI-pseu2}.

Another interesting platform for exploring the novel properties of interactions is the ultracold gases consisting of particles possessing large quadrupole moment, such as alkaline-earth and rare-earth atoms in the metastable $^{3}P_2$ states~\cite{Nagel2003, Loftus2001, Takahashi2007, Schreck2013, Takahashi2011,Schreck2010,Sterr2009, Schreck2009, Killian2009} and homonuclear diatomic molecules~\cite{Stellmer2012, Reinaudi2012, Gould2015, Schreck2017}. Following the approach for the dipolar interaction, a widely used pesudopotential for the quadrupole-quadrupole interaction (QQI) between two polarized (along the $z$ axis) bosonic quadrupoles is~\cite{Li2013,Lahrz2015,Andreev2017,Wang2018}
\begin{align}\label{QQI-pseu1}
\mathcal{V}_{\rm qq}(\mathbf r)=\frac{4\pi \hbar^2 a_{00}}{M}\delta(\mathbf r)+g_Q\frac{Y_{40}(\hat{\mathbf r})}{r^5},
\end{align}
where $M$ is the mass of the bosons, $a_{00}$ is the $s$-wave scattering length, $g_Q=\Theta^2/(\sqrt{\pi}\varepsilon_0)$ is the QQI strength with $\Theta$ being the electric quadrupole moment and $\varepsilon_0$ the vacuum permittivity, $r=|{\mathbf r}|$, and $\hat{\mathbf r}={\mathbf r}/r$. In a relevant study, Pikovski considered the general form for the anisotropic interaction~\cite{Pikovski2014}.

Armed with pseudopotential Eq.~\eqref{QQI-pseu1}, various properties of quadrupolar gases were investigated for ultracold quadrupolar Bose gases. In particular, Li {\it et al.} studied the shapes, stability, mobility, and collisions of solitons of quadrupolar gases in two-dimensional lattices~\cite{Li2013}. Lahrz {\it et al.} explored the exotic roton excitations in a two-dimensional quadrupolar Bose-Einstein condensates (BECs)~\cite{Lahrz2015}. Andreev studied the Bogoliubov spectrum of a BEC in the presence of both dipolar and quadrupolar interactions~\cite{Andreev2017}. Wang and Yi investigated the ground-state properties and stability of quadrupolar condensates by numerically solving the Gross-Pitaevskii equation~\cite{Wang2018}.

Furthermore, Bhongale {\it et al.} showed that QQI might lead to the unconventional Bardeen-Cooper-Schrieffer and charge density wave phases for quadrupolar Fermi gases trapped in a 2D square optical lattice~\cite{Bhongale2013}. Huang {\it et al.} found that quadrupolar Fermi gases in coupled one-dimensional tubes supported the triplet superfluid and spin-density wave phases~\cite{Huang2014}. More recently, using the diffusion Monte Carlo technique, Astrakharchik {\it et al.} predicted a quantum phase transition from a gas to a solid phase in a two-dimensional Bose system with quadrupolar interactions~\cite{Astrakharchik2021}.

For the experimental detection of the quadrupolar effects, Lahrz {\it et al.} proposed to measure the mean-field induced frequency shift in a two-dimensional optical square lattice of Yb or Sr atoms in the ${}^3P_2$ state~\cite{Lahrz2014}. Interestingly, Han {\it et al.} experimentally observed the quadrupolar blockade in a gas of Rb atoms~\cite{Han2022}.

Although the pseudopotential Eq.~\eqref{QQI-pseu1} is widely used, its validity has not been strictly checked through scatterings calculations~\cite{DDI-pseu1,DDI-pseu2,You1998}. In this work, we study the low-energy scattering of two identical particles interacting via a simple model potential consisting of a hard core and a bare QQI. We show that, similar to the dipolar scatterings~\cite{You1998,You2001,Blume2008}, a series of broad resonances appear on the $s$-wave scattering length as the quadrupolar interaction strength increases. Interestingly, sequences of broad resonances also appear on the generalized scattering lengths of the $p$ wave for fermions and of the $s$ and $d$ wave coupling for bosons, which is in striking contrast to dipolar scatterings. We further show that the scattering amplitudes for higher partial waves with incoming ($l$) and outgoing ($l'$) channels satisfying $l+l'\geq 4$ are determined by the first Born approximation. These observations inspire us to propose new pseudopotentials for quadrupolar interaction which incorporates the anisotropic short-range contributions due to the $p$-wave scattering for fermions and the $s$ and $d$ waves coupling for bosons. Finally, to further understand the scattering resonances, we also study the bound-state properties of two particles in both free space and harmonic traps.

The rest of this paper is organized as follows. In Sec.~\ref{secform}, we introduce the model potential and give a brief analysis to the threshold behavior of the quadrupolar scattering. In Sec.~\ref{secresu}, we present results on the generalized scattering lengths and the bound-state properties in the vicinity of collision resonances. We also propose a new pseudopotential for the quadrupolar interaction between identical bosons or fermions. Finally, we conclude in Sec.~\ref{seconcl}.

\section{Formulation}\label{secform}
We consider the collisions between two identical polarized quadrupoles. The interaction potential is modeled as
\begin{align}
V_{\rm model}({\mathbf r})=\left\{\begin{array}{ll}
g_QY_{40}(\hat {\mathbf r})/r^5, &\mbox{ for $r>r_c$},\\
\infty, &\mbox{ for $r\leq r_c$},
\end{array}\right.\label{model-intqq}
\end{align}
where we have introduced a short-distance truncation $r_c$ such that the interaction potential is a hard sphere for $r<r_c$ and a pure quadrupolar interaction for $r>r_c$. Apparently, the QQI is anisotropic as it depends on the polar angle $\theta$ of $\hat{\mathbf r}$. In particular, the QQI is repulsive along $\theta=0^\circ$ and $90^\circ$ and is most attractive along $\theta=49.1^\circ$. We point out that the quadrupolar interactions were also involved in the studies of the cold collisions between the metastable alkaline-earth atoms~\cite{QQI-scattering1,Kokoouline2003,QQI-scattering2}.

The quadrupolar scatterings are described by the Schr\"odinger equation
\begin{align}
\left[-\frac{\hbar^2\nabla^2}{2\mu}+V_{\rm model}({\mathbf r})\right]\psi({\mathbf r})=E\psi({\mathbf r}),\label{scheqsc}
\end{align}
where $\mu=M/2$ is the reduce mass of the colliding atoms and $E=\hbar^2 k^2/(2\mu)$ is the incident energy with $k=|{\mathbf k}|$ being the incident momentum. To proceed, we expand the scattering wave function in terms of the partial waves, i.e., $\psi({\mathbf r})=\sum_{lm}r^{-1}\phi_{lm}(r)Y_{lm}(\hat {\mathbf r})$, where $l$ is the orbital angular momentum quantum number and $m$ is projection quantum number. After substituting the partial wave expansion into Eq.~\eqref{scheqsc}, we obtain, for $r>r_c$, the coupled equations
\begin{align}
&\left[\frac{d^2}{dr^2}-\frac{l(l+1)}{r^2}+k^2\right]\phi_{lm}-\frac{2\bar g_Q}{r^5}\sum_{l'm'}\zeta_{lm}^{l'm'}\phi_{l'm'}=0,\label{scatering-QQI}
\end{align}
where $\bar g_Q=g_Q\mu/\hbar^2$ characterizes the quadrupolar interaction strength and is of the dimension of volume and
\begin{align}\label{eqzeta}
\zeta_{lm}^{l'm'}&=\int d\hat r Y_{lm}^*(\hat {\mathbf r})Y_{40}(\hat {\mathbf r})Y_{l'm'}(\hat {\mathbf r})=(-1)^m\nonumber\\
&\quad\times \sqrt\frac{9(2l+1)(2l'+1)}{4\pi}\begin{pmatrix}l'&4&l\\-m'&0&m\end{pmatrix}\begin{pmatrix}l'&4&l\\0&0&0\end{pmatrix}.
\end{align}
Apparently, because the model potential $V_{\rm model}$ conserves the $z$ component of the orbital angular momentum, $\zeta_{lm}^{l'm'}=0$ if $m\neq m'$. Without loss of generality, we shall restrict ourselves to the $m=0$ case as scattering channels with different $m$ quantum numbers can be treated separately. Moreover, the $3j$ symbols in Eq.~\eqref{eqzeta} imply that $\zeta_{lm}^{l'm'}$ is nonzero only when $|l-l'|\leq 4\leq l+l'$ and $l+l'$ is even. As a result, QQI does not directly contribute to the lowest partial waves due to $\zeta_{00}^{00}=\zeta_{00}^{20}=0$ for bosons and $\zeta_{10}^{10}=0$ for fermions.

At sufficiently large $r$, the scattering wave functions satisfy the asymptotical boundary conditions
\begin{align}
\phi_{l0}(r)\xrightarrow{r\to\infty}Y_{l0}^*(\hat {\mathbf k})r j_l(kr)-\sum_{l'} K_{l0}^{l'0}(k)Y_{l'0}^*(\hat{\mathbf k})r n_{l'}(kr),\label{boundary-condition}
\end{align}
where $j_l(x)$ and $n_l(x)$ are the spherical Bessel and spherical Neumann functions, respectively, and $K^{l'0}_{l0}$ is the element of the $K$ matrix corresponding to the incoming channel $(l0)$ and the outgoing channels $(l'0)$. Physically, the first term at the right-hand-side of Eq.~\eqref{boundary-condition} is the free spherical wave solution and the second one accounts the contributions of the interaction. For a spherical potential, the $K$ matrix elements reduce to the familiar form $K^{l'0}_{l0}=\delta_{ll'}\tan\delta_l(k)$ with $\delta_l(k)$ being the phase shift. While for an $1/r^n$-type anisotropic interaction, $K_{l0}^{l'0}$ behaves in the low-energy limit ($k\to0$) as $k^{l+l'+1}$ if $l+l'<n-3$ and as $k^{n-2}$ otherwise~\cite{TYWu}. Consequently, in the low-energy limit, the generalized scattering lengths are then defined as
\begin{align}\label{threshold}
a_{ll'}\equiv a_{l0}^{l'0}=\left\{
\begin{array}{ll}
\!-\displaystyle{\lim_{k\to 0} k^{-1}K^{00}_{00}},&\mbox{for $l=l'=0$},\\
\!-\displaystyle{\lim_{k\to 0} k^{-3}K^{l'0}_{l0}},&\mbox{otherwise},
\end{array}
\right.
\end{align}
where $a_{00}$ has dimension of length and all other $a_{ll'}$'s are of dimension of volume. To find the generalized scatterings, we numerically integrate Eq.~\eqref{scatering-QQI} from $r=r_c$ up to $10^4r_c$ using Johnson's log-derivative propagator method~\cite{log-Johnson}. The $K$ matrix elements can then be obtained by matching the scattering wave function with the asymptotic boundary conditions Eq.~\eqref{boundary-condition}, which subsequently leads to the generalized scattering lengths.

\section{Results}\label{secresu}
Before we present the results on the generalized scattering lengths and the bound states, let us first specify the interaction parameter covered in this work. The typical quadrupole moment of alkaline-earth atoms and homonuclear molecules is about $10-40$ a.u.~\cite{Dere2001,Loftus2002,Santra2003,Santra2004,Buch2011,Byrd2011} and, without loss of generality, we choose $r_c=100$ a.u.. Then the dimensionless quadrupolar interaction strength $\bar g_Q/r_c^3$ can be as large as 1000 for Yb atom in $^3P_2$ state ($\Theta=30$ a.u.~\cite{Buch2011}), which, as shall be shown, is sufficiently large for the experimental observations of the quadrupolar scattering effects. It can be estimated that the magnetic dipole-dipole interaction between Yb atoms is much smaller than the QQI within the interatomic distance of a few hundreds Bohr radii, a range in which the atomic collision takes place~\cite{Kokoouline2003}. We therefore neglect the magnetic dipole-dipole interaction in our calculations.

Numerically, we solve Eq.~\eqref{scheqsc} with the incident energies $E/E_{r_c}=4\times 10^{-3}$, $4\times 10^{-4}$, and $4\times 10^{-6}$, where $E_{r_c}=\hbar^2/(\mu r_{c}^2)$ is a characteristic energy associated with $r_c$. For Yb atoms, these incident energies correspond to temperatures $8\times 10^{-7}$, $8\times 10^{-8}$, and $8\times 10^{-10}\, {\rm K}$, respectively. It is found numerically that the generalized scattering lengths quickly converge as the collision energy $E$ is lowered. Finally, for practical purpose, we introduce a truncation $l_{\rm cut}$ for $l$ in numerical calculations. It turns out that, for all results presented in this work, $l_{\rm cut}=34$ for bosons and $35$ for fermions are sufficient to ensure the convergence of the scattering wave functions.

\subsection{Generalized scattering lengths}\label{genscalen}

\begin{figure}[ptb]
\centering
\includegraphics[width=0.9\columnwidth]{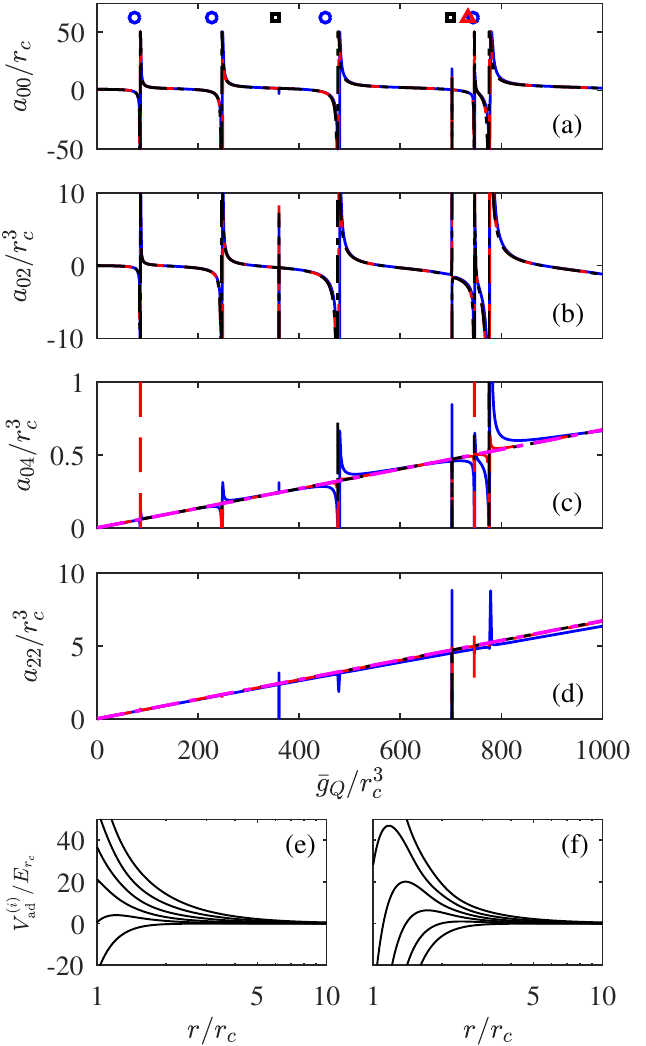}
\caption{Generalized scattering lengths $a_{00}/r_c$ (a), $a_{02}/r_c^3$ (b), $a_{04}/r_c^3$ (c), and $a_{22}/r_c^3$ (d) of identical bosons as functions of $\bar g_Q/r_c^3$. The scattering energies are $E/E_{r_c}=4\times 10^{-3}$ (blue solid line), $4\times 10^{-4}$ (red dashed line), $4\times 10^{-6}$ (black dotted line). The resonance positions shown in (a) are predicted by the WKB method using the lowest ($\Circle$), first-excited ($\square$), and second-excited ($\triangle$) adiabatic curves. The dash-dotted lines in (c) and (d) are from the Born approximation Eq.~\eqref{BA-scattering-volume} which are in excellent agreement with multi-channel numerical calculations away from resonances. (e) and (f) show the lowest 6 adiabatic curves for $\bar g_Q/r_c^3=100$ and $800$, respectively.}
\label{scabose}
\end{figure}

Let us first consider the scatterings between two identical bosons. Figure~\ref{scabose}(a) plots $a_{00}$ for two identical bosons as functions of the quadrupolar interaction strength $\bar g_Q$. As can be seen, $a_{00}$ exhibits a series of resonances as $\bar g_Q$ increases, which means that there is effective attractive potential despite of $\zeta_{00}^{00}=0$, in analogy to the situation in dipolar scatterings. The appearance of these resonances implies that zero-energy bound states continues to emerge as attractive interaction is deepened. The positions of the resonances can then be estimated using the Wentzel–Kramers–Brillouin (WKB) phase of the adiabatic potential curves~\cite{WKB1,WKB2,WKB3}, i.e.,
\begin{align}
\phi_{\rm WKB}^{(i)}(\bar g_Q)=\int_{r_c}^R dr\sqrt{-2\mu V_{\rm ad}^{(i)}(r)/\hbar^2},\label{wkbphase}
\end{align}
where $V_\mathrm{ad}^{(i)}$ is the $i$th adiabatic potential curve obtained by diagonalizing $V_\mathrm{model}+{\mathbf L}^2/(2\mu r^2)$ in the partial-wave basis. Here $\mathbf{L}$ is the total angular momentum. In Eq.~\eqref{wkbphase}, the upper bound of the integral is $R=\infty$ for the adiabatic curves without a barrier at zero collision energy; otherwise, $R$ represents the classical turning point of the barrier. In Fig.~\ref{scabose}(e) and (f), we plot the lowest six adiabatic potential curves for the bosons with $\bar g_Q/r_c^3=100$ and $800$, respectively. As can be seen, the lowest adiabatic curve is indeed attractive and becomes deepened as $\bar g_Q$ is increased. Except for the lowest adiabatic curve, all higher-lying adiabatic curves possess an energy barrier due to the centrifugal potential. A zero-energy bound state emerges whenever $\phi_{\rm WKB}^{(i)}+\pi/4$ passes through an integer multiple of $\pi$. The markers at the top of Fig.~\ref{scabose}(a) denote the positions estimated using the WKB phases. More specifically, the circles ($\Circle$), squares ($\square$), and triangle ($\triangle$) are obtained using the lowest, first-excited, and second-excited adiabatic potential curves, respectively. Following the convention, the resonances associated with the adiabatic curves without centrifugal barrier are termed potential resonances; otherwise, they are called shape resonances.

Although the resonance positions are not predicted accurately due to the anisotropy and long-range nature of QQI, the WKB phase estimation captures all the resonances for the range of interaction strength covered in the figure. In addition, it helps to understand the origin of the broad and narrow resonances. Namely, the adiabatic curves without a centrifugal barrier lead to broad resonances; while the higher-lying adiabatic curves induce narrow shape resonances as the atoms must tunnel through the centrifugal barrier.

We then turn to consider the generalized scattering lengths for the higher partial waves. In general, because, away from the shape resonances, the slowly moving particles can hardly tunnel through the centrifugal barrier, the scattering wave function for higher partial waves are essentially undisturbed by the interaction potential. As a result, the generalized scattering lengths are mainly determined by the first Born approximation, just like what has been seen in dipolar scattering. According to the Born approximation, the $K$ matrix elements can be expressed as
\begin{align}\label{BA}
\widetilde K^{l'0}_{l0}=-2\bar g_{Q}\zeta^{l'0}_{l0}k^{3}\int_{kr_c}^\infty \frac{d(kr)}{(kr)^3} j_{l'}(kr) j_l(kr).
\end{align}
Then in the low-energy limit ($k\rightarrow 0$), the generalized scattering lengths in the first Born approximation become
\begin{align}
\tilde a_{ll'}=-\lim_{k\rightarrow0}k^{-3}\widetilde K_{l0}^{l'0}=2\zeta^{l'0}_{l0}\chi_{ll'}\bar g_Q,\label{BA-scattering-volume}
\end{align}
where
\begin{align}
\chi_{ll'}=\frac{48 (-1)^{(l-l')/2} (l+l'-4)!! (l-l' -5)!!}{(l+l'+4)!! (l-l' +3)!!}.
\end{align}
Apparently, the first Born approximation gives rise to a linear dependence of the generalized scattering lengths on the interaction strength.

\begin{figure}[ptb]
\centering
\includegraphics[width=0.9\columnwidth]{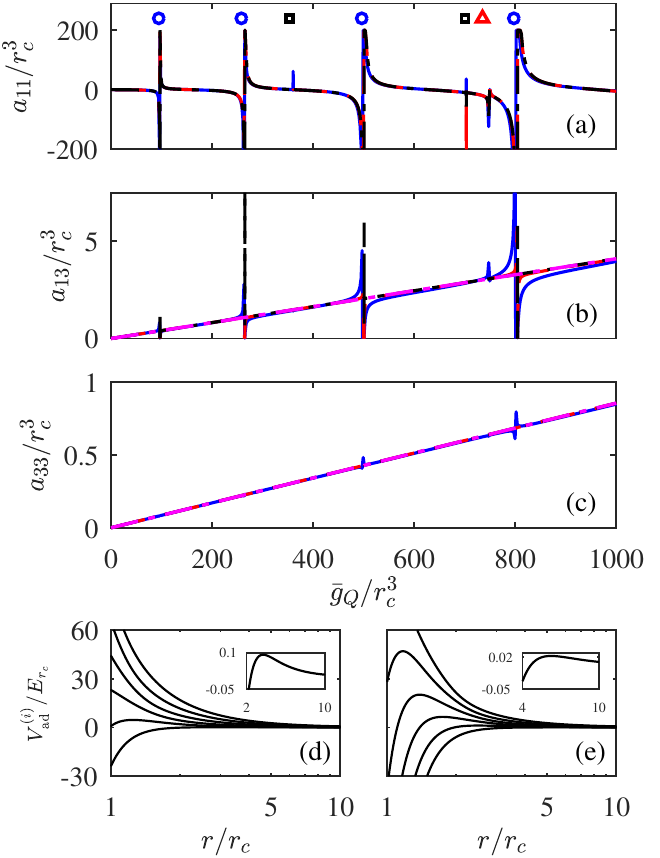}
\caption{Generalized scattering lengths $a_{11}/r_c^3$ (a), $a_{13}/r_c^3$ (b), and $a_{33}/r_c^3$ (c) of identical fermions as functions of $\bar g_Q/r_c^3$. The scattering energies are $E/E_{r_c}=4\times 10^{-3}$ (blue solid line), $4\times 10^{-4}$ (red dashed line), $4\times 10^{-6}$ (black dotted line). The resonance positions shown in (a) are predicted by the WKB method using the lowest ($\Circle$), first-excited ($\square$), and second-excited ($\triangle$) adiabatic curves. The dash-dotted lines in (b) and (c) are from the Born approximation Eq.~\eqref{BA-scattering-volume} which are in excellent agreement with multi-channel numerical calculations away from resonances. (d) and (e) show the lowest 6 adiabatic curves for $\bar g_Q/r_c^3=100$ and $800$, respectively. The insets are the zoom-in plots for the lowest adiabatic curves on which the centrifugal barriers appear.}
\label{scafermi}
\end{figure}

In Fig.~\ref{scabose}(b)-(d), we present the $\bar g_Q$ dependence of the generalized scattering lengths $a_{ll'}$ for higher partial waves. Let us first examine $a_{04}$ and $a_{22}$, for which we also plot the corresponding Born approximation results, i.e., Eq.~\eqref{BA-scattering-volume}, in Fig.~\ref{scabose}(c) and (d), respectively. As can be seen, away from scattering resonances, both $a_{04}$ and $a_{22}$ are well described by the first Born approximation. In fact, this observation holds true for all partial waves satisfying $l+l'\geq 4$. We note that since the particles are hardly scattered by the short-range potential, the first Born approximation is contributed by the long-range part of the interaction. On the other hand, $a_{02}$, as shown in Fig.~\ref{scabose}(b), exhibits many resonances as $\bar g_Q$ grows, in analogy to $a_{00}$. Moreover, the resonance positions of $a_{02}$ are identical to those of $a_{00}$'s. The underlying reason is that $\zeta_{00}^{20}=0$ such that the first Born approximation, or, equivalently, the long-range part of the interaction, does not contribute to $a_{02}$. This behavior is in striking contrast to the dipolar scattering.

The generalized scattering lengths for quadrupolar scatterings of two identical fermions are summarized in Fig.~\ref{scafermi}. Here $a_{11}$ shows a sequence of resonances as $\bar g_Q$ increases, in contrast to the dipolar scatterings where $a_{11}$ is mainly determined by the Born approximation. This result again is due to the vanishing $\zeta_{10}^{10}=0$ such that $a_{11}$ is mostly determined by the short-range part of the interaction. It is worthwhile to mention that, as shown in Fig.~\ref{scafermi}(d) and (e), although the lowest adiabatic curve always has a energy barrier, the width of the induced resonances are still very broad, compared to those induced by the higher-lying adiabatic curves. Therefore, we shall refer to the resonances induced by the lowest adiabatic curve as the broad resonances and those by higher-lying adiabatic curves as narrow resonances. Finally, as shown in Fig.~\ref{scafermi}(b) and (c), the generalized scattering length $a_{ll'}$ for higher waves ($l+l'\geq 4$) are mainly determined by the Born approximation, similar to the case for bosons.

\subsection{Pseudopotentials for quadrupolar interactions}
From the scattering calculations, it is clear that the simple pseudopotential, Eq.~\eqref{QQI-pseu1}, for quadrupolar interaction is inappropriate as the contribution of $a_{02}$ is missed. Here we construct a new quadrupolar pseudopotential by following the approach of Huang and Yang~\cite{pseu} and Derevianko~\cite{DDI-pseu3}. To this end, we first note that the regularized zero-range pseudopotential can be generally expressed as
\begin{align}
\hat{\mathcal{V}}^{({\rm reg})}=\sum_{ll'm}\hat v_{lm}^{l'm},\label{pesudopot}
\end{align}
where $l$ and $l'$ are even (odd) for bosons (fermions) and $\hat v_{lm}^{l'm}$ are operators defined by their action on an arbitrary $\mathbf r$-dependent wave function $\psi(\mathbf r)$~\cite{DDI-pseu3}, i.e.,
\begin{align}
\hat v_{lm}^{l'm}\psi(\mathbf r)&=g_{lm}^{l'm}\frac{4\pi\delta(\mathbf r)}{r^{l'}}Y_{l'm}(\hat{\mathbf r})\nonumber\\
&\quad \times\left[\frac{\partial^{2l+1}}{\partial r^{2l+1}}r^{l+1}\int Y^*_{lm}(\hat{\mathbf r})\psi(\mathbf r)d\hat{\mathbf r}\right]_{r\to0}
\end{align}
with the coupling coefficients $g_{lm}^{l'm}$ being defined as
\begin{align}
g_{lm}^{l'm}=-\frac{\hbar^2}{M}\frac{K_{lm}^{l'm}}{k^{l+l'+1}}\frac{(2l+1)!!(2l'+1)!!}{(2l+1)!}.\nonumber
\end{align}
Since, away from resonances, the $K$ matrix elements, $K_{lm}^{l'm'}$, with $l+l'\geq 4$ originate from the Born approximation, their contributions are completely covered by the bare quadrupolar interaction $\bar g_QY_{40}(\hat{\mathbf r})/r^5$. Then, based on the scattering calculations, we only need to take care of the $K_{00}^{00}$, $K_{00}^{02}$, and $K_{02}^{00}$ terms for bosons and the $K_{1m}^{1m}$ terms for fermions in the pseudopotential Eq.~\eqref{pesudopot}. Consequently, in the low-energy limit, the pseudopotential for identical bosons and fermions can be straightforwardly written out as
\begin{align}
\hat{\mathcal{V}}_{\rm qq}^{(B)}\psi(\mathbf r)&=\frac{4\pi\hbar^2 a_{00}}{M}\delta(\mathbf r)\frac{\partial}{\partial r}(r\psi)+\frac{\sqrt{4\pi}\hbar^2 a_{02}}{M}\delta(\mathbf r)\nonumber\\
&\quad\times\left[\frac{60\pi}{r^2}Y_{20}(\hat{\mathbf r})\frac{\partial}{\partial r}(r \psi)+\frac{1}{8}\frac{\partial^5}{\partial r^{5}}r^{3}\!\!\int\!\! d\hat{\mathbf r} Y_{20}(\hat{\mathbf r})\psi \right]\nonumber\\
&\quad+g_Q\frac{Y_{40}(\hat{\mathbf{r}})}{r^5}\psi\label{QQI-pseu2}
\end{align}
and
\begin{align}
\hat{\mathcal{V}}_{\rm qq}^{(F)}\psi(\mathbf r)&=\sum_{m}\frac{6\pi\hbar^2a_{1m}^{1m}}{M}\frac{\delta(\mathbf r)}{r}Y_{1m}(\hat{\mathbf r})\nonumber\\
&\quad\times\left[\frac{\partial^3}{\partial r^3}r^2 \int d\hat{\mathbf r}Y^*_{1m}(\hat{\mathbf r})\psi\right]+g_Q\frac{Y_{40}(\hat{\mathbf{r}})}{r^5}\psi,\label{QQI-pseu-fermi}
\end{align}
respectively. Here the long-range feature of the QQI is captured by bare quadrupolar interaction and the short-range characteristic is accounted by the $a_{00}$ and $a_{02}$ for bosons and by $a_{1m}$ for fermions.

\subsection{Two-body bound states}
To gain more insight into the collision resonances, we explore the bound-state properties of two quadrupoles in both free space and trapping potentials. In addition, the energy spectrum of two quadrupolar fermionic atoms in a harmonic trap can be used to calculate the virial coefficient, a quantity that determines the high-temperature thermodynamics of strongly interacting gases~\cite{Peng2011}.

The free-space bound state problem is described by Eq.~\eqref{scheqsc} with a negative energy $E$; while the trapped case is governed by the equation
\begin{align}
\left[-\frac{\hbar^2\nabla^2}{2\mu}+\frac{1}{2}\mu\omega^2 r^2+V_{\rm model}(\mathbf r)\right]\psi=E\psi,\label{schtrap}
\end{align}
where $\omega$ is the trap frequency. In both cases, we numerically solve the Schr\"odinger equations using $B$-splines (see, e.g., Ref.~\cite{B-spline}). In particular, we shall focus on the bound states in the vicinity of collision resonances.

\begin{figure}[ptb]
\centering
\includegraphics[width=\columnwidth]{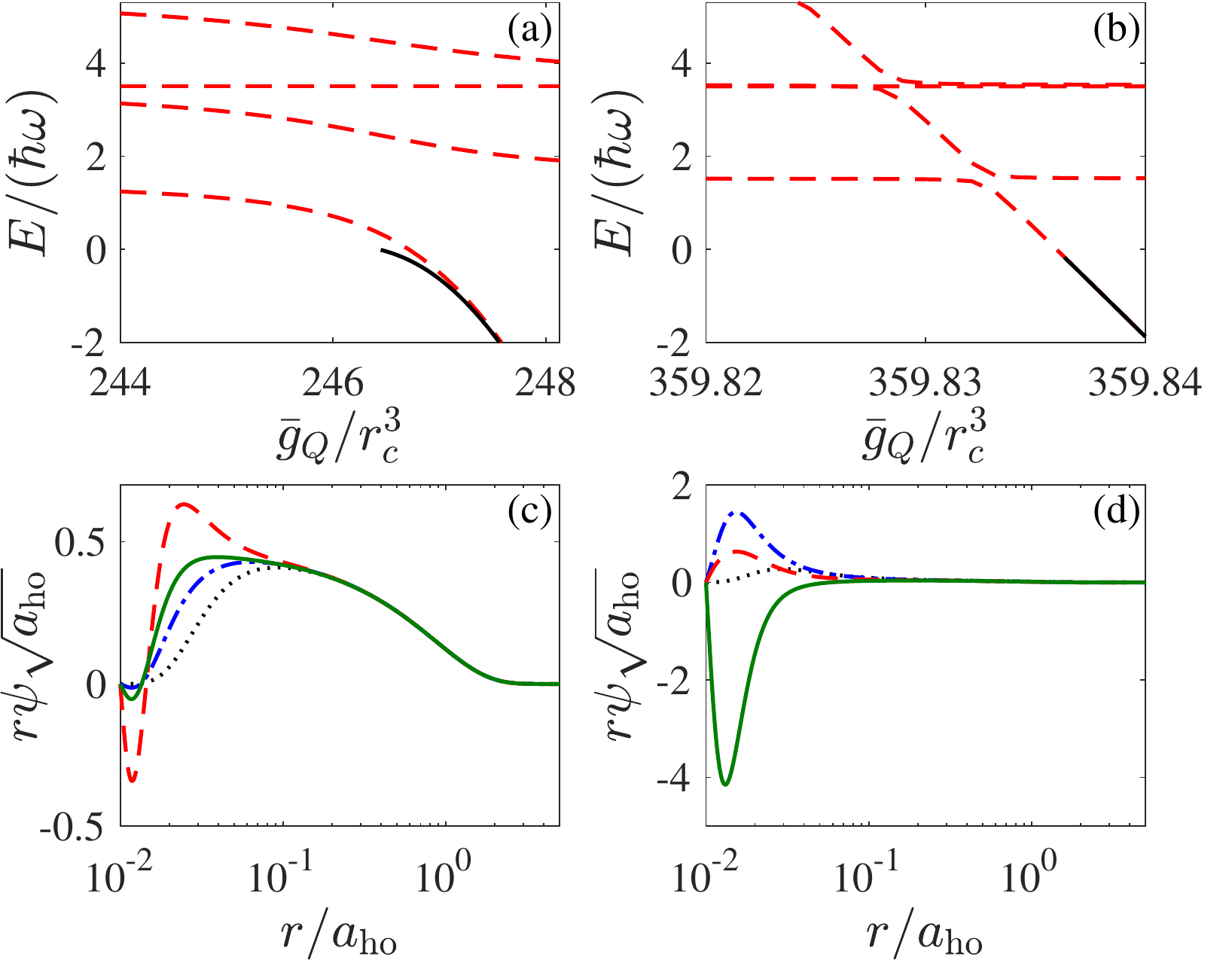}
\caption{Bound-state structures of two identical bosons around the resonances at $\bar g_Q/r_c^3=247.0$ (left panels) and $359.84$ (right panels). (a) and (b) plot the QQI strength dependence of the bound-state energies in free space (solid lines) and trap (dashed lines). (c) and (d) are the bound-state wave function of trapped bosons with energy $E=-0.5\hbar\omega$ and $\theta=0^\circ$ (dotted line), $30^\circ$ (dash-dotted lines), $49.1^\circ$ (dashed lines), and $60^\circ$ (solid line). Here $a_{\rm ho}=\sqrt{\hbar/(\mu\omega)}$ is the oscillator length and the size of the hard core is $r_c=0.01a_{\rm ho}$.}
\label{energy-spectrum-boson}
\end{figure}

Figure~\ref{energy-spectrum-boson} summarizes the main results for the bound-state properties of two identical bosons. The dashed lines in Fig.~\ref{energy-spectrum-boson}(a) and (b) plot the eigenenergies of trapped bosons as function of $\bar g_Q$ around a broad and a narrow resonance, respectively. In the vicinity a resonance, the lowest energy level with positive energy quickly drops and becomes negative as $\bar g_Q$ increases, signaling that the two atoms initially bounded by harmonic potential become a deeply bound molecule state bounded by the QQI. Also, in Fig.~\ref{energy-spectrum-boson}(a) and (b), the solid lines below zero energy in these figures represent the corresponding eigenenergies of the bound states in free space, which are in very good agreement with the eigenenergies of the trapped systems as long as the bound energy is sufficiently large. This observation suggests that the effects of the confining potential is negligible for the molecule-like states.

\begin{figure}[ptb]
\centering
\includegraphics[width=\columnwidth]{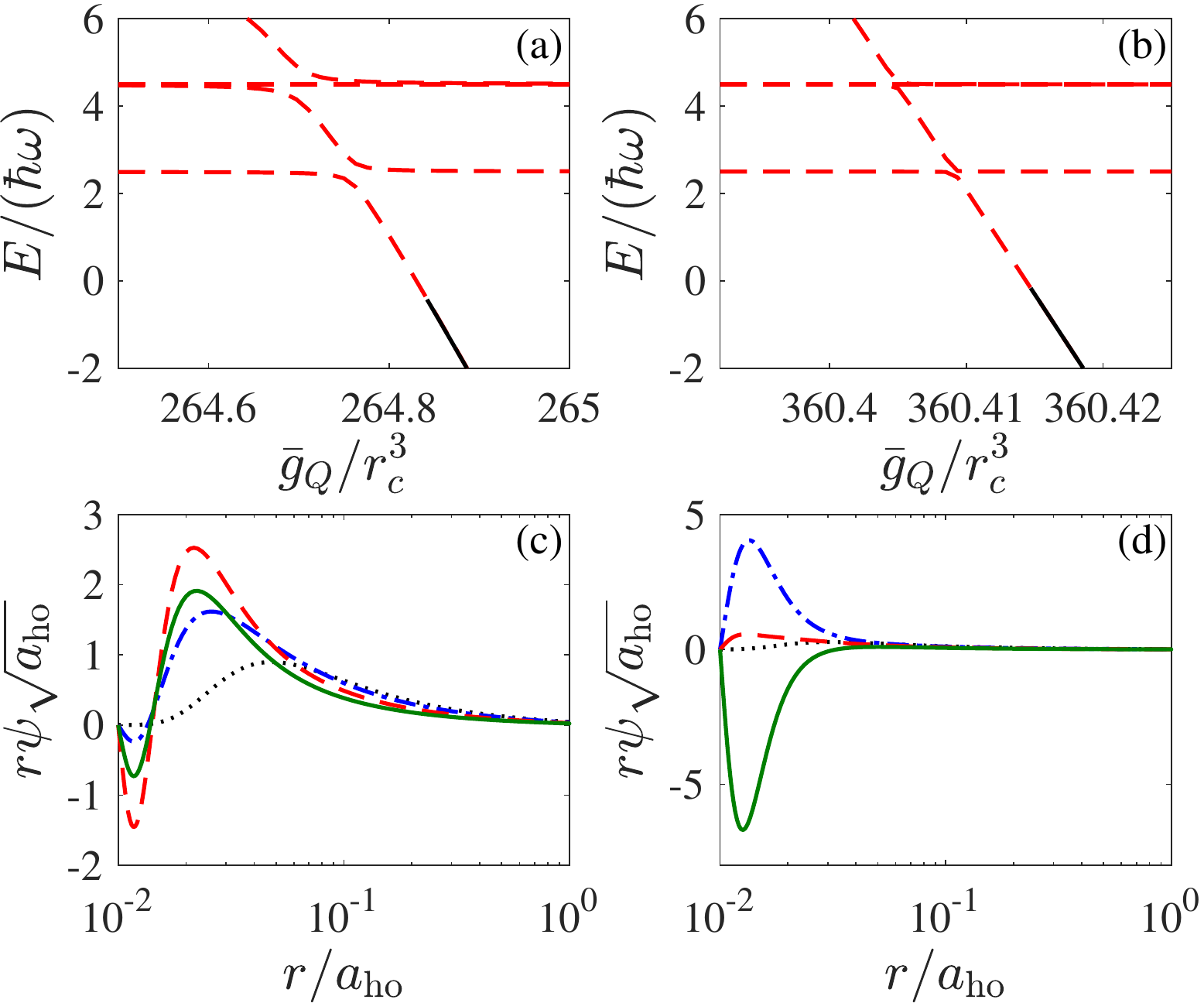}
\caption{Bound-state structures of two identical fermions around the resonances at $\bar g_Q/r_c^3=264.84$ (left panels) and $360.42$ (right panels). (a) and (b) plot the QQI strength dependence of the bound-state energies in free space (solid lines) and trap (dashed lines). (c) and (d) are the bound-state wave function of trapped fermions with energy $E=-0.5\hbar\omega$ and $\theta=0^\circ$ (dotted line), $30^\circ$ (dash-dotted lines), $49.1^\circ$ (dashed lines), and $60^\circ$ (solid line).}
\label{energy-spectrum-fermion}
\end{figure}

As to the bound-state wave functions, Fig.~\ref{energy-spectrum-boson}(c) shows the rescaled radial wave function, $r\psi$, of the bound states corresponding to the eigenenergy $E=-0.5\hbar\omega$ in the vicinity of the broad resonance at $\bar g_Q/r_c^3=247.0$. The amplitude of the wave function is largest along the most attractive direction $\theta=49.1^\circ$ and is considerably lower along the repulsive directions, in particular, along $\theta=0^\circ$. Still this wave function is dominated by the isotropic $l=0$ wave ($\sim99\%$). This observation remains true for the wave function of a bound state around other broad resonance. On the other hand, the bound-state wave function around a narrow resonance is dramatically different from that around a broad one. As shown in Fig.~\ref{energy-spectrum-boson}(d), this wave function mainly consists of the three partial waves: $l=2$ ($34\%$), $6$ ($47\%$), and $10$ ($9.6\%$). It should be noted that, in the vicinity of a narrow resonance, the partial-wave composition may vary rapidly as the interaction strength changes. In addition, the bound-state wave function around a narrow resonance is more localized at the center than that around a broad one, which is related to the large centrifugal barrier associated with higher partial waves.

Finally, for the bound states of two identical fermions, Fig.~\ref{energy-spectrum-fermion}(a) and (b) show the QQI dependence of the eigenenergies around a broad ($\bar g_Q/r_c^3=264.84$) and a narrow ($\bar g_Q/r_c^3=360.42$) resonances, respectively. It can be seen that the eigenenergies structure is very similar to that of the bosonic bound states. Moreover, as shown in Fig.~\ref{energy-spectrum-fermion}(c), the wave function around a broad resonance is dominantly contributed by the $p$ wave (over $88\%$). While around a narrow resonance [see Fig.~\ref{energy-spectrum-fermion}(d)], the main contributions to the wave function come from a wide range of partial waves, including $l=3$ ($31\%$), $5$ ($33\%$), $7$ ($10\%$), and $9$ ($17\%$), which is again similar to its bosonic counterpart.

\section{Conclusion}\label{seconcl}
In summary, we have studied the low-energy scattering and the bound-state properties of two identical particles interacting via QQI. We numerically compute the generalized scattering lengths $a_{ll'}$ as functions of the quadrupolar interaction strength. It has been shown that the short-range part of the interaction potential gives rise to $a_{00}$ and $a_{02}$ for bosons and $a_{1m}$ for fermions. These generalized scattering lengths exhibit a series of scattering resonances as the quadrupolar interaction strength grows. On the other hand, the long-range part of the interaction, i.e., the bare QQI, contributes to the generalized scattering lengths through the first Born approximation. Consequently, we propose new pseudopotentials that correctly take into account the contributions of $a_{02}$ and $a_{1m}$ for bosonic and fermionic quadrupoles, respectively. These pseudopotentials should pave the way for studying the many-body physics of the quadrupolar quantum gases. Finally, for a better understanding the scattering resonances, we have also presented a detailed analysis of the bound-state properties in the vicinity of the resonances.

\begin{acknowledgments}
We thank Tao Shi for fruitful discussions. This work was supported by the NSFC (Grants No. 12135018 and No. 12047503), by NKRDPC (Grant No. 2021YFA0718304), by NSAF (Grant No. U1930201), and by the Strategic Priority Research Program of CAS (Grant No. XDB28000000).
\end{acknowledgments}

\end{document}